\documentclass[10pt,conference]{IEEEtran}

\usepackage[T1]{fontenc}
\usepackage[utf8]{inputenc}
\usepackage{amsmath,amssymb,amsfonts}
\usepackage{algorithmic}
\usepackage{algorithm}
\usepackage{graphicx}
\usepackage{booktabs}
\usepackage{tabularx}
\usepackage{multirow}
\usepackage{array}
\usepackage{xcolor}
\usepackage{cite}
\usepackage{textcomp}
\usepackage{microtype}

\usepackage[
  colorlinks = true,
  linkcolor  = blue,
  citecolor  = blue,
  urlcolor   = blue,
  pdftitle   = {Few-Shot Cross-Device Transfer for Quantum Noise Modeling on Real Hardware},
  pdfauthor  = {Sahil Al Farib}
]{hyperref}
\usepackage{url}

\graphicspath{{figures/}}

\begin{document}

\title{Few-Shot Cross-Device Transfer for Quantum Noise Modeling on Real Hardware}

\author{%
  \IEEEauthorblockN{Sahil Al Farib}
  \IEEEauthorblockA{
    United International University\\
    Dhaka, Bangladesh\\
    \href{mailto:sfarib222186@bscse.uiu.ac.bd}{sfarib222186@bscse.uiu.ac.bd}
  }
  \and
  \IEEEauthorblockN{Sheikh Redwanul Islam}
  \IEEEauthorblockA{
    United International University\\
    Dhaka, Bangladesh\\
    \href{mailto:sislam222142@bscse.uiu.ac.bd}{sislam222142@bscse.uiu.ac.bd}
  }
  \and
  \IEEEauthorblockN{Azizur Rahman Anik}
  \IEEEauthorblockA{
    United International University\\
    Dhaka, Bangladesh\\
    \href{mailto:azizur@cse.uiu.ac.bd}{azizur@cse.uiu.ac.bd}
  }
}

\maketitle

\begin{abstract}
In the noisy intermediate-scale quantum (NISQ) regime, quantum devices contain hardware-specific noise sources which restrict device-invariant error mitigation strategies. We explore transfer learning approaches to apply noise models learned on one quantum device to a different device with the help of a small amount of data. We create a real-hardware dataset from two IBM quantum devices, \texttt{ibm\_fez} (source) and \texttt{ibm\_marrakesh} (target), comprising 170 noisy and ideal circuit output distributions, with device calibration features added.

We train a residual neural network on the source device to map noisy to ideal outcomes. The zero-shot transfer test shows a KL divergence of 1.6706 (up from 0.3014), establishing device specificity. With $K = 20$ fine-tuning samples, KL drops to 1.1924 (28.6\% improvement over zero-shot), recovering 34.9\% of the gap between zero-shot and in-domain KL.

Ablation studies reveal that the major cause of mismatches across devices is CX gate error, followed by readout error. The results show quantum noise can be learned and fine-tuned with minimal samples, and provide a plausible approach to cross-device quantum error mitigation.
\end{abstract}


\section{Introduction}

The noisy intermediate-scale quantum (NISQ) era of quantum computing is constrained by the hardware noise caused by decoherence, gate and measurement error~\cite{preskill2018nisq}. Such noise patterns are not generic and vary across devices, over time (due to calibration drift) and across qubit registers of the same physical chip~\cite{dasgupta2021stability}. This makes it challenging to develop noise models transferable across different devices.

Data-driven approaches to quantum error mitigation have seen growing interest. Methods such as zero-noise extrapolation (ZNE)~\cite{li2017efficient}, probabilistic error cancellation (PEC)~\cite{temme2017error}, and Clifford data regression~\cite{czarnik2021error} can reduce the effect of noise on expectation values. However, these techniques typically require per-device calibration data or assume that noise characteristics are stable and transferable—an assumption our results directly challenge. More recent machine learning approaches have modeled noise using circuit structure and calibration features~\cite{du2026noise}, but cross-device generalization remains largely unaddressed.

The challenge of adapting a pre-trained model to a new domain with minimal data is well studied in classical machine learning under the names transfer learning and few-shot learning~\cite{pan2010survey,finn2017maml}. We ask whether this paradigm applies to quantum noise: can a model trained to predict ideal circuit outputs from noisy measurements on one IBM quantum device adapt to a different device using only a small number of target-device samples?

We propose a data-driven framework for few-shot cross-device noise adaptation and make the following concrete contributions:
\begin{enumerate}
  \item We construct a real-hardware paired dataset of 170 (noisy, ideal) circuit output distributions across two IBM quantum devices (\texttt{ibm\_fez} and \texttt{ibm\_marrakesh}), paired with device calibration features spanning four hardware parameters.
  \item We train a residual neural network on the source device and demonstrate that zero-shot transfer to the target device increases KL divergence by 5.5$\times$ relative to the in-domain baseline, confirming that noise profiles are strongly device-specific.
  \item We show that few-shot fine-tuning with $K = 5, 10, 20$ target-device samples yields monotonic improvement: at $K = 20$, KL divergence falls from 1.6706 to 1.1924—a 28.6\% reduction relative to the zero-shot baseline, or equivalently, a 34.9\% recovery of the gap between zero-shot and in-domain performance (gap $= 1.6706 - 0.3014 = 1.3692$; recovered $= 0.4782$).
  \item We perform a calibration drift analysis and a leave-one-out feature ablation study, identifying CX gate error as the strongest cross-device mismatch feature and readout error as a secondary mismatch signal.
\end{enumerate}

\section{Related Work}

\subsection{Quantum Error Mitigation}

The limitations of quantum computing in the NISQ (noisy intermediate-scale quantum) era stem from hardware noise such as decoherence, imperfect gates and measurements~\cite{preskill2018nisq}. Various quantum error mitigation (QEM) strategies have been suggested to reduce their impact without achieving fault tolerance.

Zero-noise extrapolation (ZNE) introduces artificial noise and extrapolates back to zero noise to offer hardware-independent error mitigation~\cite{li2017efficient}. Probabilistic error cancellation (PEC), on the other hand, learns an inverse noise model via quasi-probability sampling, allowing for unbiased results with a higher sample cost~\cite{temme2017error}. Clifford data regression also uses classically simulable circuits to learn correction functions from noisy to ideal outputs, showing impressive results on single-device quantum computers~\cite{czarnik2021error}.

Recently, machine learning methods have shown potential. These approaches leverage classical machine learning models to model the noise process or directly learn the correction from noisy to ideal outcomes, reducing the need for detailed physical models of noise~\cite{du2026noise,strikis2021learning}. However, current methods either assume a single device or need to be retrained for each physical implementation, making them challenging to scale across diverse quantum devices.

\subsection{Device Variability and Noise Drift}

A primary difficulty in quantum computing is that noise processes are not universal and change over time due to calibration drift. Recent empirical research has demonstrated that even seemingly identical quantum devices can have vastly different noise characteristics, which affects the portability of models~\cite{dasgupta2021stability}.

This suggests that noise mitigation techniques acquired on one device may not perform well on another. This poses a problem for many current approaches, which rely on device-specific calibration or repeated characterization, resulting in significant additional cost, and preventing their use in practical quantum workflows.

\subsection{Transfer Learning and Few-Shot Adaptation}

The challenge of transferring knowledge from one domain to another with scarce data is a well-studied problem in classical machine learning (ML) under the paradigms of transfer learning and few-shot learning. Transfer learning allows the knowledge acquired from one domain to be used to enhance learning in another domain~\cite{pan2010survey}.

Meta-learning methods like Model-Agnostic Meta-Learning (MAML) also build on this concept, learning initializations of model parameters that can quickly adapt to new tasks with few training samples~\cite{finn2017maml}. These approaches have seen remarkable success in vision, language and science applications, but in the context of quantum noise modeling, they have yet to be extensively explored.

\subsection{Gap in Existing Work}

While there have been considerable advances in both quantum error mitigation (QEM) and transfer learning, the combination of these techniques is still in its infancy. Current QEM methods typically aim to enhance performance on a single device, while classical transfer learning methods have not yet been rigorously investigated for adapting to different quantum noise patterns.

Specifically, there is a lack of empirical evidence that indicates whether patterns of noise learned from one device can be transferred to another (or quickly adapted) using a few additional data points. This is essential for realizing scalable, data-driven quantum error mitigation that can be applied across different devices.

\section{Dataset and Experimental Setup}

\subsection{Circuit Generation}

We generate a controlled set of 85 quantum circuits spanning four structural families to ensure coverage of diverse noise regimes (Table~\ref{tab:circuit_types}).

\begin{table}[htbp]
  \centering
  \caption{Circuit Type Distribution}
  \label{tab:circuit_types}
  \begin{tabular}{lcp{3.5cm}}
    \toprule
    \textbf{Circuit Type} & \textbf{Count} & \textbf{Primary Purpose} \\
    \midrule
    Random circuits    & 40 & Structural diversity; generalization \\
    Bell state circuits & 15 & Two-qubit entanglement; CX error sensitivity \\
    GHZ state circuits & 15 & Multi-qubit entanglement; error accumulation \\
    QFT circuits       & 15 & Layered gate accumulation; coherent errors \\
    \bottomrule
  \end{tabular}
\end{table}

All circuits are parameterized with 2--5 qubits and circuit depth 2--8. Circuit generation uses a fixed random seed (\texttt{seed = 42}) for full reproducibility.

\begin{figure}[htbp]
  \centering
  \includegraphics[width=\linewidth]{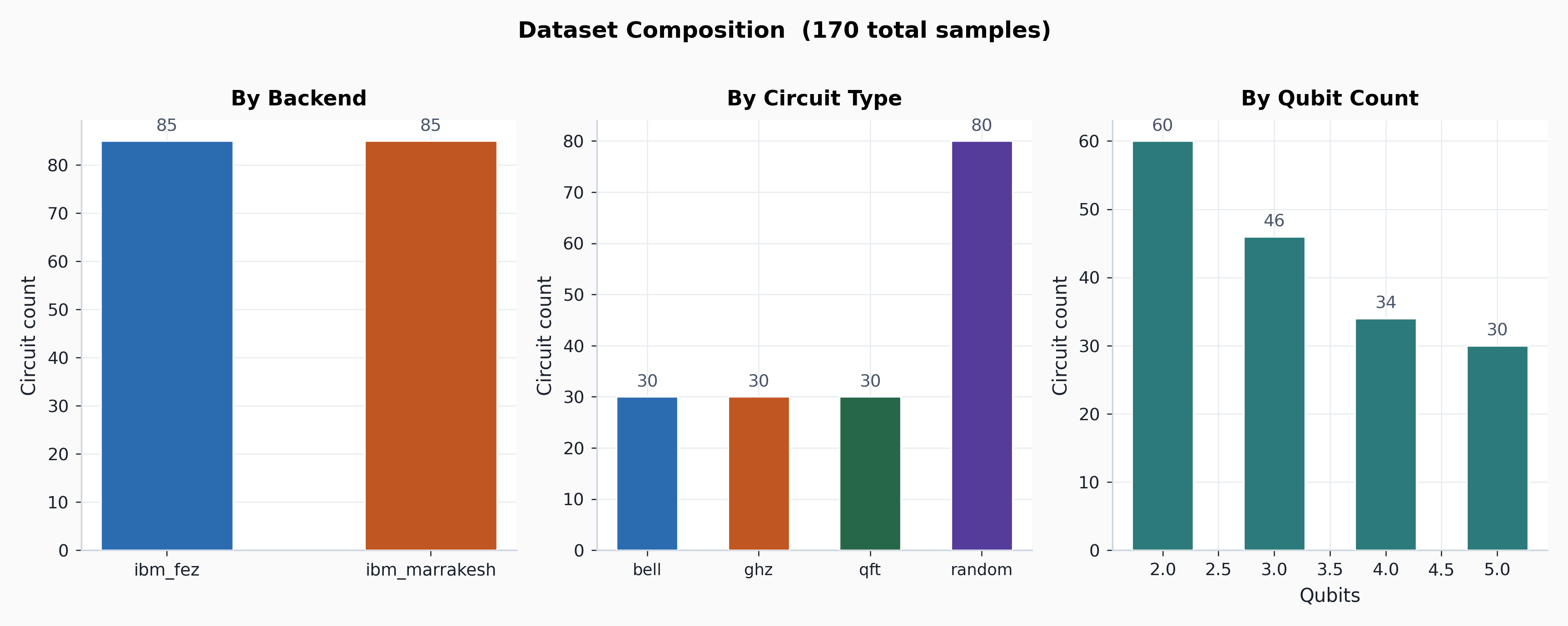}
  \caption{Dataset composition. (a)~Sample counts by backend: 85 per device. (b)~Circuit type distribution: 40 random, 15 Bell, 15 GHZ, 15 QFT. (c)~Qubit count distribution across 2--5 qubits. The balanced structure across circuit types ensures the dataset covers a representative range of noise regimes.}
  \label{fig:dataset}
\end{figure}

\subsection{Data Collection}

For each circuit we collect three data sources:

\textbf{Noisy output distribution.} Each circuit is executed on real IBM quantum hardware via the Qiskit IBM Runtime service. All circuits for a given backend are submitted as a single batch job to minimize calibration drift within a backend session. Empirical probability distributions are computed from hardware measurement outcomes.

\textbf{Ideal output distribution.} Each circuit is simulated using Qiskit's noiseless statevector simulator, giving the ground-truth ideal probability distribution.

\textbf{Calibration features.} Device calibration data is retrieved from the IBM Quantum calibration API at the time of circuit execution, capturing: mean $T_1$ (qubit relaxation time), mean $T_2$ (qubit dephasing time), mean readout error, and mean CX gate error, averaged across all active qubits.

We run all 85 circuits on two IBM quantum backends:
\begin{itemize}
  \item \textbf{Backend A — \texttt{ibm\_fez}} (source device): used exclusively for model training.
  \item \textbf{Backend B — \texttt{ibm\_marrakesh}} (target device): reserved for zero-shot and few-shot evaluation.
\end{itemize}
The final dataset contains 170 paired samples (85 per backend).

We do not control for temporal drift between backend executions; instead, we treat each backend snapshot as representative of a distinct device state, which is consistent with practical usage of cloud-based quantum hardware.

\subsection{Feature Representation}

Each sample is encoded as a fixed-length feature vector $\mathbf{x} \in \mathbb{R}^{41}$.

The first 9 dimensions are scalar features (Table~\ref{tab:features}):

\begin{table}[htbp]
  \centering
  \caption{Input Feature Vector (Indices 0--8)}
  \label{tab:features}
  \begin{tabular}{cl}
    \toprule
    \textbf{Index} & \textbf{Feature} \\
    \midrule
    0 & Number of qubits \\
    1 & Circuit depth \\
    2 & CX gate count \\
    3 & H gate count \\
    4 & X gate count \\
    5 & Mean $T_1$ (standardized) \\
    6 & Mean $T_2$ (standardized) \\
    7 & Mean readout error (standardized) \\
    8 & Mean CX gate error (standardized) \\
    \bottomrule
  \end{tabular}
\end{table}

Dimensions 9--40 are the noisy output distribution over computational basis states, padded to a fixed size of $32 = 2^5$ (supporting circuits up to 5 qubits; unused entries are zero-padded).

The 9 scalar features are standardized to zero mean and unit variance using statistics computed from Backend A training data only. Backend B samples are standardized with the same Backend A statistics at inference time, so calibration feature differences between devices remain interpretable by the model.

\textbf{Target} $\mathbf{y} \in \mathbb{R}^{32}$ is the ideal output distribution over computational basis states, padded to size 32.

\begin{figure}[htbp]
  \centering
  \includegraphics[width=\linewidth]{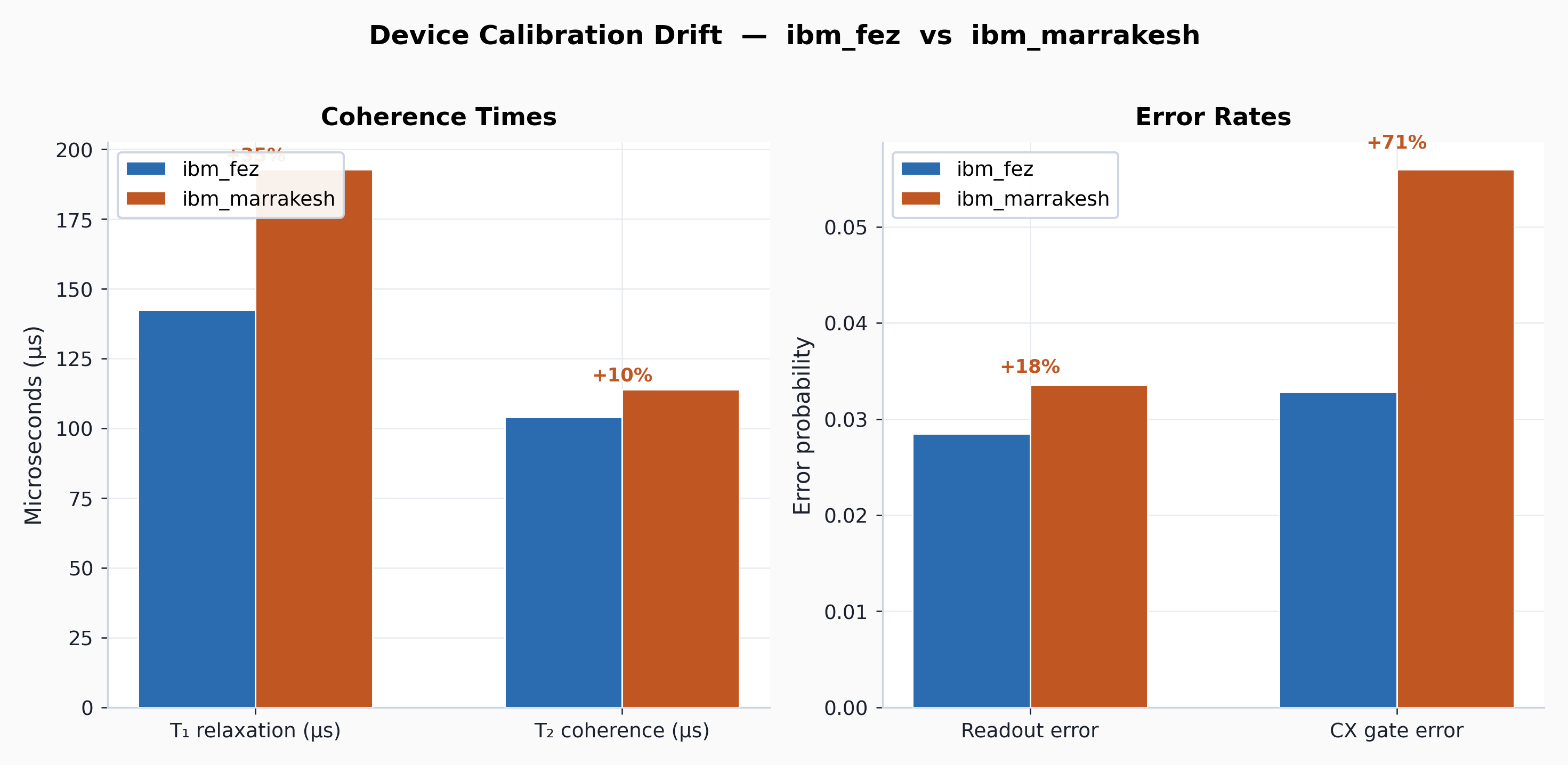}
  \caption{Device calibration drift between \texttt{ibm\_fez} (Backend A) and \texttt{ibm\_marrakesh} (Backend B). Left: coherence times $T_1$ and $T_2$ in microseconds. Right: readout error and CX gate error rates. Backend B has longer coherence times but higher gate and readout error rates—a qualitatively distinct noise profile that motivates the need for explicit adaptation.}
  \label{fig:calibration_drift}
\end{figure}

\section{Method}

\subsection{Problem Formulation}

We model noise mitigation as supervised distribution regression. Given a feature vector $\mathbf{x}$ encoding the noisy output distribution, circuit structure, and device calibration, we learn:
\begin{equation}
  f_\theta : \mathbf{x} \;\longrightarrow\; \hat{\mathbf{y}} \approx \mathbf{y}_{\text{ideal}}
\end{equation}
Training minimizes KL divergence between the predicted and true ideal distributions:
\begin{equation}
  \mathcal{L}(\theta) = D_{\mathrm{KL}}\!\left(\mathbf{y}_{\text{ideal}} \,\Big\|\, \mathrm{softmax}\!\left(f_\theta(\mathbf{x})\right)\right)
\end{equation}
computed via PyTorch's \texttt{F.kl\_div} with \texttt{reduction="batchmean"}. 

Given a source device Ds and a target device Dt, we study whether a model trained on Ds can be adapted to Dt using K labeled samples.

\subsection{Model Architecture: Residual Noise Adapter}

We design a Residual Noise Adapter (RNA), a multilayer perceptron that learns a residual correction over the noisy input distribution rather than predicting the ideal distribution from scratch:
\begin{equation}
  \hat{\mathbf{y}} = \mathrm{softmax}\!\left(\mathbf{x}_{\text{noisy}} + f_\theta(\mathbf{x})\right)
\end{equation}
where $\mathbf{x}_{\text{noisy}} = \mathbf{x}[9{:}] \in \mathbb{R}^{32}$ is the noisy distribution sub-vector and $f_\theta(\mathbf{x}) \in \mathbb{R}^{32}$ is the learned per-state correction. The backbone $f_\theta$ consists of:
\begin{align*}
  &\text{Linear}(41 \to 128) \to \text{LayerNorm}(128) \to \text{GELU} \to \text{Dropout}(0.10) \\
  &\text{Linear}(128 \to 128) \to \text{LayerNorm}(128) \to \text{GELU} \to \text{Dropout}(0.10) \\
  &\text{Linear}(128 \to 64) \to \text{LayerNorm}(64) \to \text{GELU} \\
  &\text{Linear}(64 \to 32)\quad\text{[head]}
\end{align*}

Model hyperparameters are summarized in Table~\ref{tab:hyperparams}.

\begin{table}[htbp]
  \centering
  \caption{Model Hyperparameters}
  \label{tab:hyperparams}
  \begin{tabular}{ll}
    \toprule
    \textbf{Parameter} & \textbf{Value} \\
    \midrule
    Input dimension      & 41 \\
    Output dimension     & 32 \\
    Hidden dimensions    & $128 \to 128 \to 64$ \\
    Activation           & GELU \\
    Normalization        & LayerNorm after each hidden layer \\
    Dropout              & 0.10 (after first two blocks) \\
    Head                 & Linear($64 \to 32$), no activation \\
    Residual connection  & adds $\mathbf{x}[9{:}]$ to head output before softmax \\
    \bottomrule
  \end{tabular}
\end{table}

The residual formulation provides two key benefits. First, it encourages near-identity initialization: when the backbone weights are small, the correction $f_\theta(\mathbf{x}) \approx \mathbf{0}$ and the model outputs approximately the noisy input—a more stable starting point than an arbitrary mapping. Second, it constrains the model to learn a correction over the existing noisy distribution, an inductive bias that aligns with how physical noise distorts ideal distributions.

\subsection{Training Protocol}

The model is trained exclusively on \textbf{Backend A (\texttt{ibm\_fez})} data.

\textbf{Data split.} Backend A's 85 samples are split 80/20 into training (68 samples) and validation (17 samples) using a fixed generator seed (\texttt{seed = 42}).

\textbf{Optimization settings} are summarized in Table~\ref{tab:optimization}. The best model checkpoint is saved based on minimum validation KL divergence.

\begin{table}[htbp]
  \centering
  \caption{Optimization Settings}
  \label{tab:optimization}
  \begin{tabular}{ll}
    \toprule
    \textbf{Setting} & \textbf{Value} \\
    \midrule
    Optimizer              & AdamW \\
    Learning rate          & $1 \times 10^{-3}$ \\
    Weight decay           & $1 \times 10^{-4}$ \\
    LR scheduler           & ReduceLROnPlateau (factor=0.5, patience=12) \\
    Batch size             & 16 (train), 32 (validation) \\
    Max epochs             & 250 \\
    Early stopping patience & 25 epochs (on validation KL) \\
    \bottomrule
  \end{tabular}
\end{table}

\begin{figure}[htbp]
  \centering
  \includegraphics[width=\linewidth]{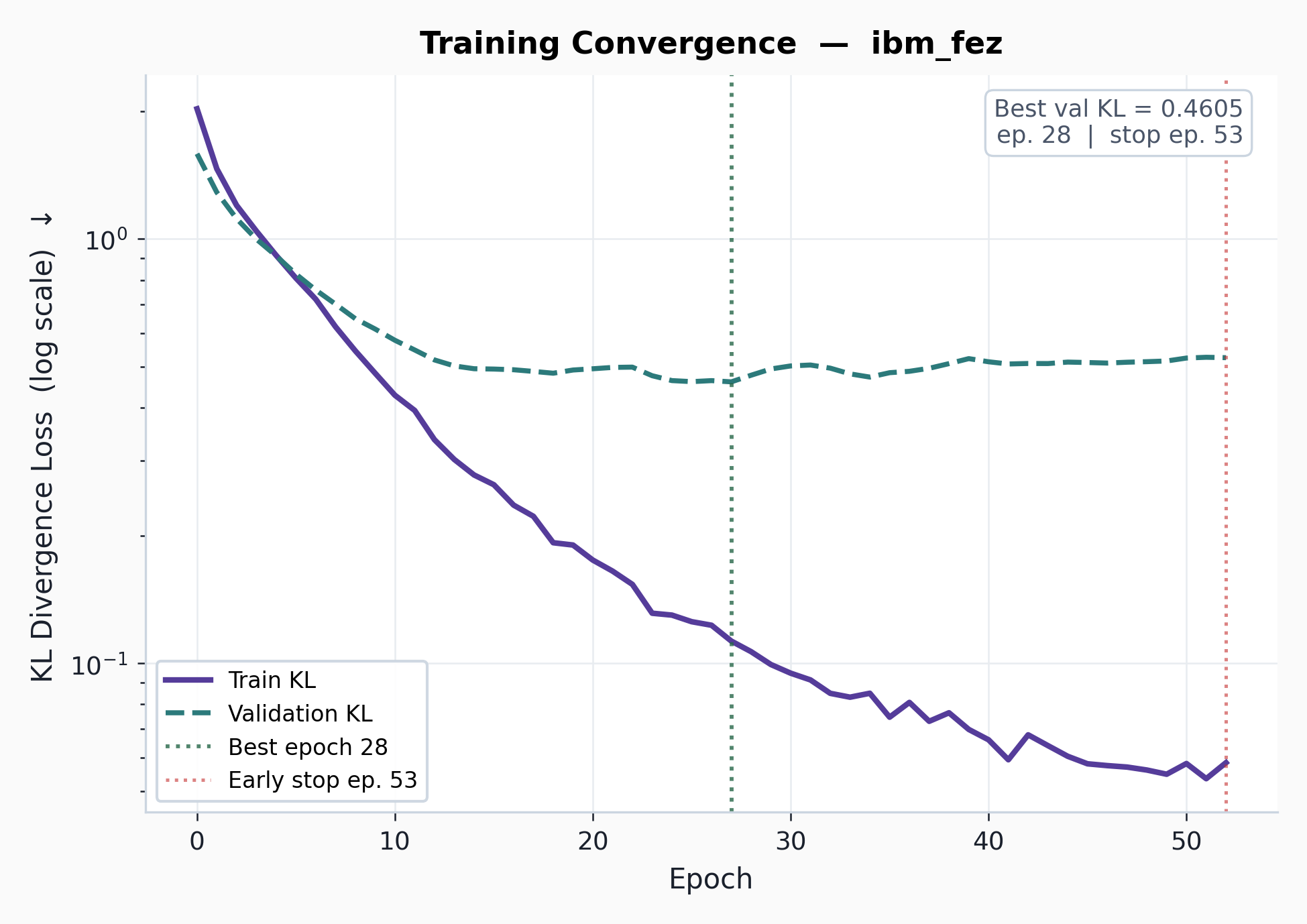}
  \caption{Training convergence curve (KL divergence, log scale) on Backend A. Both training and validation loss decrease monotonically before early stopping. The gap between training and validation curves is small, indicating limited overfitting despite the small dataset size (68 training samples). Best val KL $= 0.4605$ at epoch 27; early stopping at epoch 52.}
  \label{fig:training}
\end{figure}

\subsection{Few-Shot Adaptation Protocol}

After training on Backend A, we evaluate the model in three settings:

\textbf{In-domain (A$\to$A).} The model is evaluated on a held-out split of Backend A data (same train/val split, \texttt{seed = 42}). This serves as an upper-bound reference for achievable performance.

\textbf{Zero-shot (A$\to$B).} The trained Backend A model is applied directly to all 85 Backend B samples without any modification. This quantifies the cost of cross-device noise mismatch.

\textbf{Few-shot (A$\to$B, $K$ shots).} The pre-trained model is fine-tuned on $K$ randomly sampled circuits from Backend B, then evaluated on the remaining $(85 - K)$ Backend B samples.

The adaptation procedure is detailed in Table~\ref{tab:adaptation}.

\begin{table}[htbp]
  \centering
  \caption{Few-Shot Adaptation Settings}
  \label{tab:adaptation}
  \resizebox{\columnwidth}{!}{%
  \begin{tabular}{lll}
    \toprule
    \textbf{Setting}            & $K \le 10$                         & $K = 20$ \\
    \midrule
    Trainable layers            & Head only                           & Last hidden block $+$ head \\
    Learning rate               & $1 \times 10^{-4}$                  & $5 \times 10^{-5}$ \\
    Max epochs                  & 60                                  & 80 \\
    Early stopping patience     & 12 epochs                           & 12 epochs \\
    Replay buffer               & \multicolumn{2}{l}{24 randomly sampled Backend A training samples} \\
    \bottomrule
  \end{tabular}
  }
\end{table}

\textbf{Replay.} 24 Backend A training samples are mixed into each fine-tuning batch alongside the $K$ target samples. This prevents catastrophic forgetting of source-domain structure while adapting to the target device.

\textbf{Layer freezing strategy.} For small $K$ ($\le 10$), only the output head is updated, limiting the degrees of freedom to avoid overfitting. For $K = 20$, the final hidden block (Linear $128 \to 64$, LayerNorm, GELU) and head are unfrozen, allowing deeper adaptation while the backbone's earlier feature representations remain fixed.

Results are averaged over 5 random seeds (seeds 0--4) controlling the $K$-sample selection.

\subsection{Why Residual Distribution Learning for Quantum Noise}

In our approach, rather than directly learning to regress from features to the desired output, we make the design decision to learn the residual of the probability distributions. This is driven by both statistical and physical properties of quantum noise processes.

Statistically, the noisy output distribution already holds a strong component of the ideal distribution. The majority of quantum noise channels (such as amplitude damping, depolarizing noise and readout errors) correspond to structured distortions. Hence, training a residual correction function $f_{\theta}(x)$ enables the model to learn to correct the noisy input distribution instead of learning the entire distribution over again. This simplifies the learning task, making it more sample-efficient, an essential feature in few-shot learning.

From a physical standpoint, many noise processes exhibit approximately additive or redistributive behavior in the probability simplex. For example, readout errors tend to redistribute probability mass between neighboring bitstrings, while gate errors introduce systematic distortions correlated with circuit structure. The residual formulation

\begin{equation}
\hat{y} = \mathrm{softmax}\left(x_{\text{noisy}} + f_{\theta}(x)\right)
\end{equation}

encodes an inductive bias that the ideal distribution is a corrected version of the observed noisy distribution, rather than an unrelated target.

This approach also leads to a stable training landscape. When the network is first initialized, the network weights are small, $f_{\theta}(x) \approx 0$, and the model prediction is close to the noisy input. This prevents instability that may be observed when predicting entire probability distributions without some prior knowledge of the output structure. During training, the model gradually learns adjustments to redistribute probability mass towards the target distribution.

For transfer learning across devices, the residual formulation is beneficial. Patterns of noise (e.g., entanglement-induced distortions) are shared between devices, but magnitudes are device-dependent. The residual network can leverage structure learned on the source device and adapt only the magnitude and direction of adjustments for few-shot transfer learning. This is why it is effective to freeze the early layers and update just the final layers or "head": early layers learn device-invariant patterns while later layers learn device-specific calibration effects.

To conclude, residual distribution learning offers a principled and efficient way of modeling quantum noise, leading to effective performance within and across devices even with limited data.

\section{Results}

\subsection{Main Results}

Table~\ref{tab:results} reports cross-device transfer performance in terms of KL divergence and Total Variation distance.

\begin{table}[htbp]
  \centering
  \caption{Cross-Device Transfer Performance (KL divergence and Total Variation distance). Best few-shot result in \textbf{bold}.}
  \label{tab:results}
  \resizebox{\columnwidth}{!}{%
  \begin{tabular}{lccc}
    \toprule
    \textbf{Condition} & \textbf{KL Div.}$\downarrow$ & \textbf{TV Dist.}$\downarrow$ & \textbf{KL Improv.} \\
    \midrule
    In-domain (A$\to$A)   & 0.3014                    & 0.5023                    & — \\
    Zero-shot (A$\to$B)   & 1.6706                    & 0.5282                    & baseline \\
    Few-shot $K{=}5$      & $1.5874\!\pm\!0.0849$     & $0.5287\!\pm\!0.0081$     & $-5.0\%$ \\
    Few-shot $K{=}10$     & $1.5172\!\pm\!0.0669$     & $0.5290\!\pm\!0.0088$     & $-9.2\%$ \\
    Few-shot $K{=}20$     & $\mathbf{1.1924\!\pm\!0.0630}$ & $0.5503\!\pm\!0.0095$ & $\mathbf{-28.6\%}$ \\
    \bottomrule
  \end{tabular}
  }
\end{table}

The zero-shot KL of 1.6706 represents a 5.5$\times$ degradation relative to the in-domain baseline of 0.3014. This confirms that the noise structure of \texttt{ibm\_marrakesh} is fundamentally different from that of \texttt{ibm\_fez}, even though both are IBM superconducting processors of the same generation.

Few-shot adaptation recovers a substantial fraction of this gap. The improvement is monotonically increasing with $K$: 5.0\% at $K{=}5$, 9.2\% at $K{=}10$, and 28.6\% at $K{=}20$—all measured as percentage reduction relative to the zero-shot baseline. Expressed as recovery of the gap to in-domain performance (gap $= 1.6706 - 0.3014 = 1.3692$), $K{=}20$ recovers $(1.6706 - 1.1924)/1.3692 = \mathbf{34.9\%}$ of the total transfer deficit. The standard deviations (over 5 random seeds) remain stable across $K$ values, indicating robustness to the particular choice of $K$ calibration samples.

\begin{figure}[htbp]
  \centering
  \includegraphics[width=\linewidth]{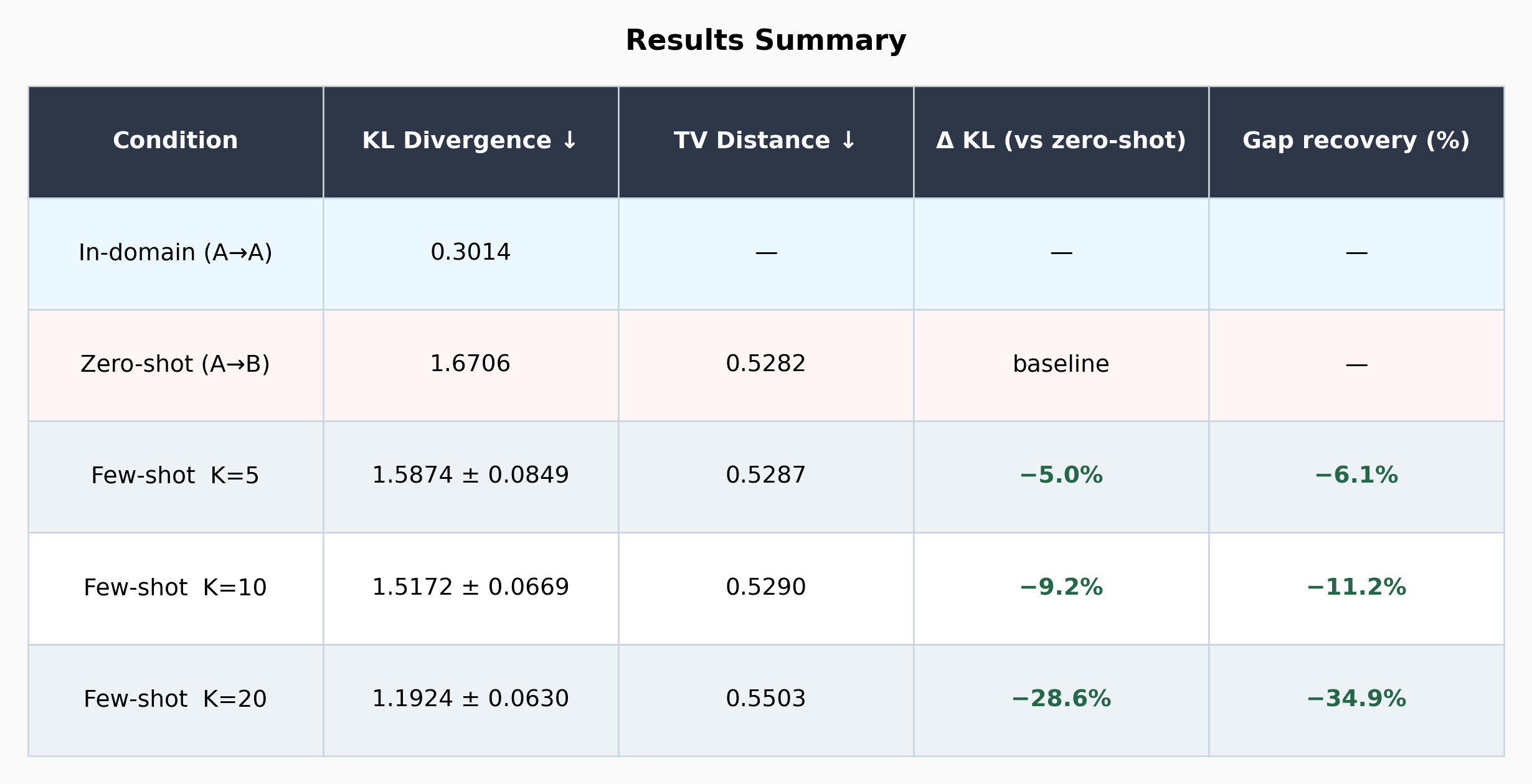}
  \caption{Results summary table. Rows: in-domain, zero-shot, and $K{=}5/10/20$ few-shot conditions. Columns: KL divergence (mean $\pm$ std), TV distance, and KL improvement percentage. Note that TV increases slightly ($+0.0221$), indicating that improvements in low-probability states (KL-sensitive) may come at the cost of small deviations in dominant states.}
  \label{fig:results_table}
\end{figure}

\begin{figure}[htbp]
  \centering
  \includegraphics[width=\linewidth]{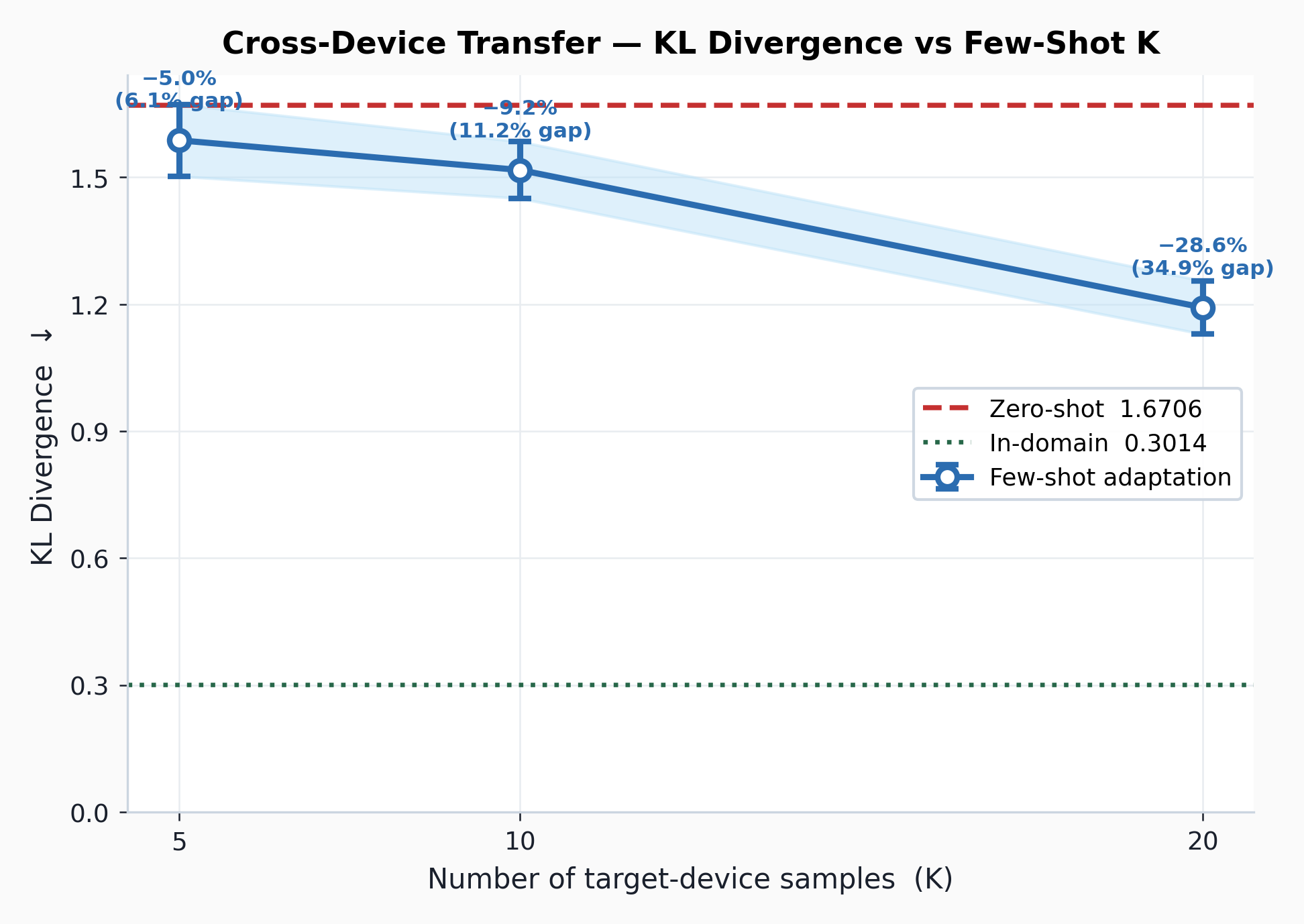}
  \caption{KL divergence as a function of few-shot sample count $K$. Dashed red line: zero-shot baseline (1.6706). Dotted green line: in-domain reference (0.3014). Blue curve with shaded $\pm1$ std confidence band: few-shot fine-tuning over 5 seeds. Percentage annotations indicate per-$K$ improvement over zero-shot. The monotonic improvement with $K$ confirms that the pre-trained source model provides a useful initialization for adaptation.}
  \label{fig:kl_transfer}
\end{figure}

\begin{figure}[htbp]
  \centering
  \includegraphics[width=\linewidth]{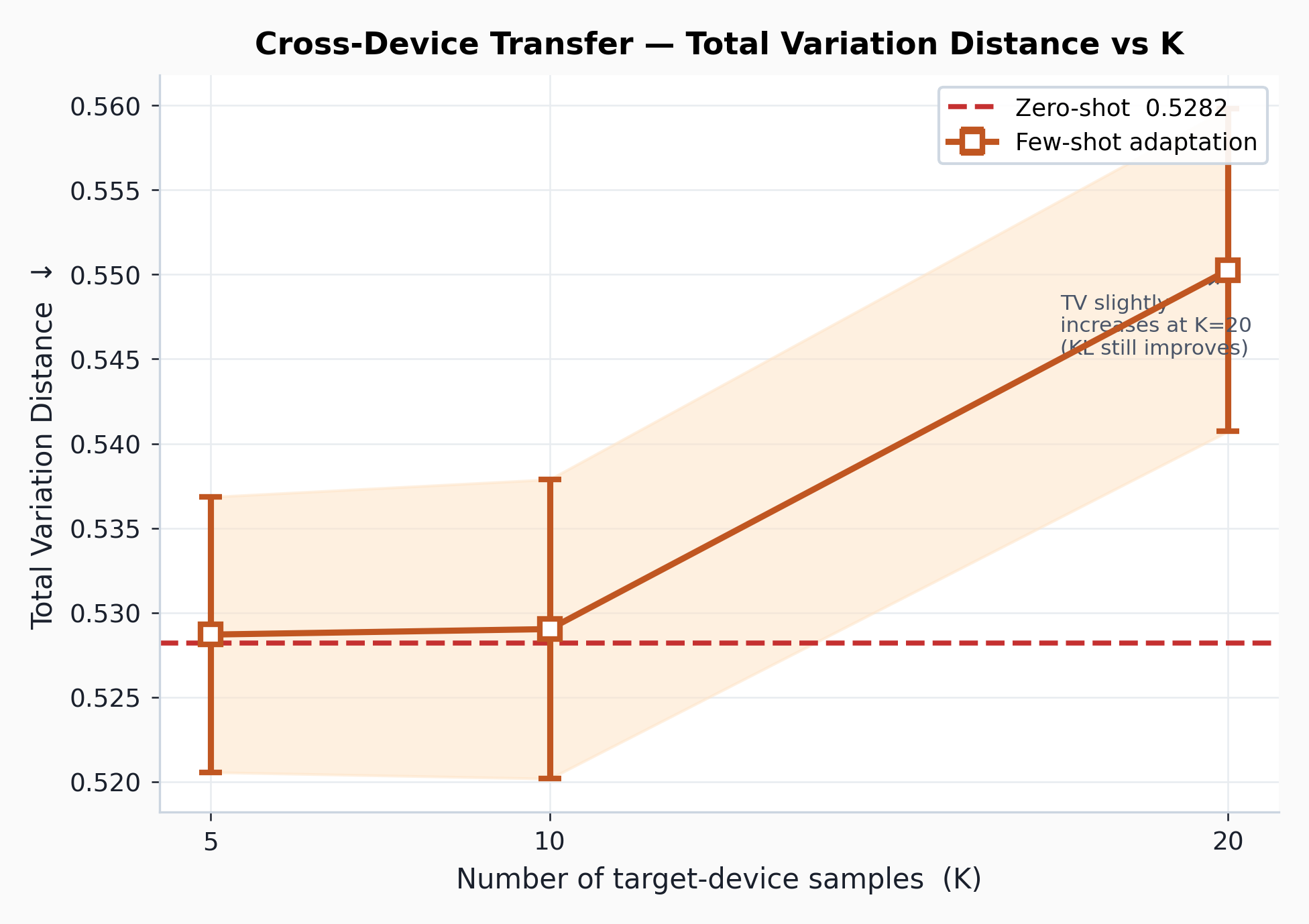}
  \caption{Total Variation distance as a function of $K$. Unlike KL divergence, TV distance does not decrease monotonically: it remains approximately flat for $K \le 10$ and increases slightly at $K = 20$ ($0.5282 \to 0.5503$). This divergence from the KL trend is discussed in Section~\ref{sec:results_main}.}
  \label{fig:tv_transfer}
\end{figure}

\label{sec:results_main}
KL divergence is more sensitive to low-probability states, while TV distance weights all states uniformly. At $K = 20$, improvements in low-probability regions reduce KL, while small deviations in dominant states lead to a slight increase in TV. This demonstrates that KL and TV capture complementary aspects of distributional alignment.

\subsection{Example Prediction}

To qualitatively verify model behavior, we select the Backend B circuit with the highest noisy-to-ideal KL divergence (the hardest example in the test set) and compare the noisy input, ideal target, and model prediction distributions.

\begin{figure}[htbp]
  \centering
  \includegraphics[width=\linewidth]{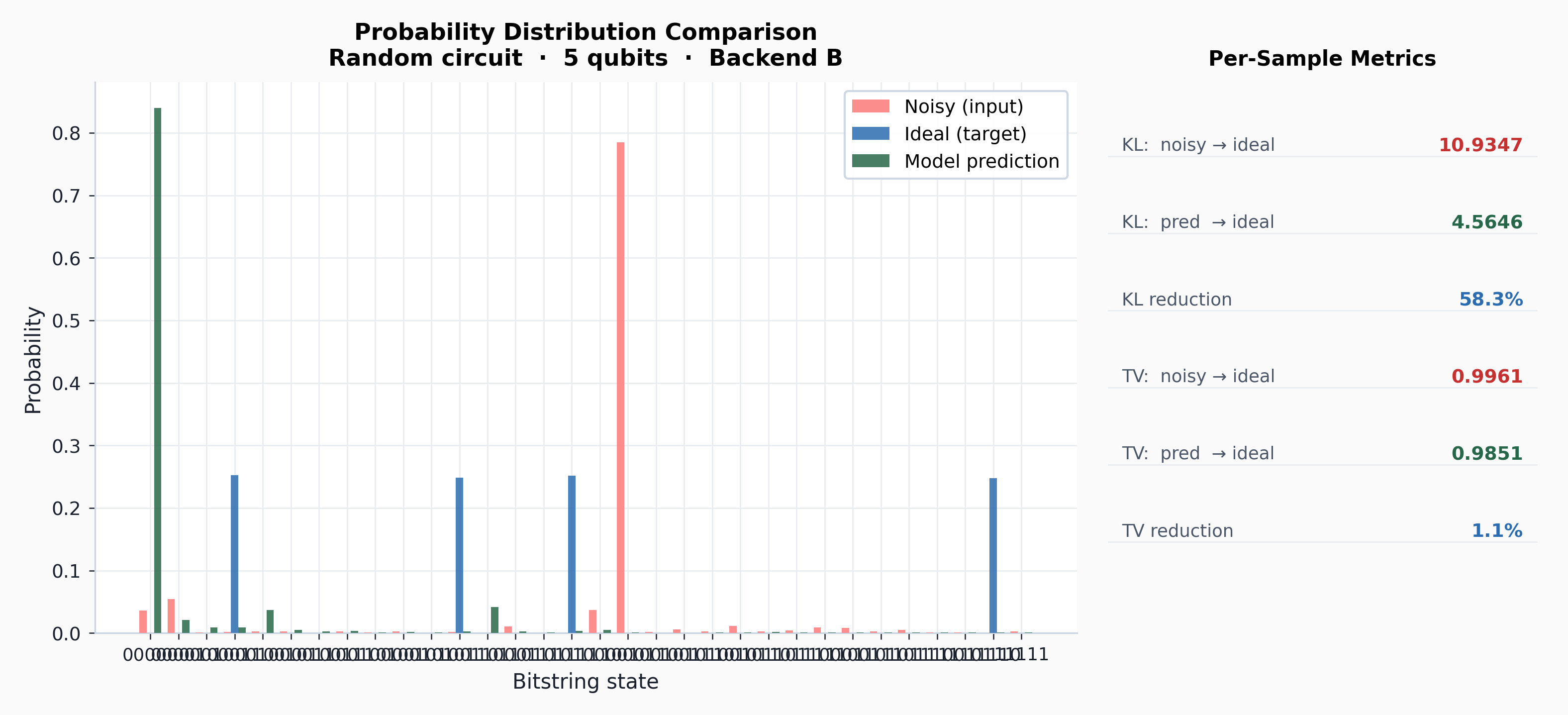}
  \caption{Probability distribution comparison for a representative circuit on Backend B (circuit type and qubit count shown in panel title). Grouped bar chart: red = noisy input, blue = ideal target, green = model prediction over computational basis states. Side panel: KL and TV metrics for each sample, for all three pairs. The model effectively redistributes mass to the correct distribution, bringing down the noisy-to-ideal KL.}
  \label{fig:example_pred}
\end{figure}

\section{Calibration Drift Analysis}

A necessary condition for cross-device transfer to be difficult is that the two devices have measurably different noise profiles. Table~\ref{tab:calibration} reports the mean calibration properties of each backend, extracted from the IBM Quantum calibration API at the time of circuit execution.

\begin{table}[htbp]
  \centering
  \caption{Device Calibration Comparison ($T_1$ and $T_2$ in microseconds; errors are dimensionless probabilities)}
  \label{tab:calibration}
  \resizebox{\columnwidth}{!}{%
  \begin{tabular}{lccc}
    \toprule
    \textbf{Property} & \textbf{ibm\_fez (A)} & \textbf{ibm\_marrakesh (B)} & $\Delta$ (B$-$A) \\
    \midrule
    $T_1$ ($\mu$s)    & 142.4 & 192.8 & $+50.5$ ($+35.4\%$) \\
    $T_2$ ($\mu$s)    & 104.1 & 114.0 & $+10.0$ ($+9.6\%$) \\
    Readout error     & 0.0285 & 0.0335 & $+0.0050$ ($+17.5\%$) \\
    CX gate error     & 0.0328 & 0.0560 & $+0.0232$ ($+70.7\%$) \\
    \bottomrule
  \end{tabular}
  }
\end{table}

The two devices present a counterintuitive pattern: Backend B has longer coherence times ($T_1$: $+35.4\%$, $T_2$: $+9.6\%$) but higher error rates (readout: $+17.5\%$, CX gate: $+70.7\%$). This indicates that the devices have qualitatively distinct noise profiles—not merely a uniform scaling of the same noise structure. A model trained on Backend A must learn a fundamentally different error correction to succeed on Backend B. The 70.7\% difference in CX gate error is particularly striking, and it directly explains the large zero-shot KL divergence (1.6706 vs.\ 0.3014 in-domain).

\section{Error Analysis: When and Why Does the Model Fail?}

Despite the effectiveness of few-shot adaptation for cross-device prediction, it is essential to identify and understand the model's limitations. We examine errors in predictions for different circuits and different distributions to understand the systematic shortcomings and inform future research.

First, prediction error is systematically higher for highly multi-qubit entangled circuits, such as GHZ and Quantum Fourier Transform (QFT) circuits. These circuits magnify the impact of multi-qubit gate errors, particularly errors on the controlled-NOT (CX) gates, which are shown to have the largest drift across devices in our calibration analysis. In these instances, the mapping from calibration features to distributional corrections learned from the training data may be partially mismatched to the noise regime of the target device, and cannot be fully corrected by few-shot learning.

Second, the model is less effective at predicting distributions with strong peaks (high-probability states) and long tails (low-probability states). The Kullback--Leibler (KL) divergence metric is heavily weighted by low-probability regions, and thus a reduction in KL divergence can be achieved despite a slight decrease in accuracy in high-probability states. This accounts for the observed discrepancy between KL and total variation (TV) distance metrics: although the KL error usually decreases as the number of adaptation samples $K$ increases, TV distance can sometimes increase due to slight changes in high-probability events.

Third, zero-shot transfer errors demonstrate that summary features of device calibrations don't entirely capture device noise characteristics. Although average $T_1$ and $T_2$ times, readout errors, and CX gate errors offer a succinct summary of the device, they exclude spatial and qubit-to-qubit interactions. This is especially the case for circuits that map onto different sets of qubits on different devices, where noise characteristics are highly variable.

Fourth, low-shot adaptation is dependent on the variability of the adaptation data. If the adaptation set lacks diversity in circuit structures (for instance, contains mainly circuits with low entanglement), the learned corrections may not be applicable to new circuits. Random seed averaging helps to smooth out variance in the results, but variability is still an issue in the extremely low-shot setting.

Despite these failure modes, the model shows improvement for all adaptation sizes, suggesting that the learned representation generalises well for device adaptation. These failure modes can be mitigated by future work through the inclusion of qubit-specific calibration data, circuit structure encodings and more sophisticated model architectures such as graph neural networks or attention mechanisms.

\section{Feature Ablation Study}

To identify which calibration features drive cross-device adaptation, we perform a leave-one-out ablation study. Each calibration scalar (feature indices 5--8 in the input vector) is zeroed out individually, and the zero-shot KL divergence on Backend B is re-evaluated. Only the input features are modified; the model weights remain unchanged.

\begin{table}[htbp]
  \centering
  \caption{Calibration Feature Ablation (baseline KL = zero-shot KL = 1.6706)}
  \label{tab:ablation}
  \begin{tabular}{lcc}
    \toprule
    \textbf{Ablation Condition} & \textbf{KL Div.} & $\Delta$ vs.\ Baseline \\
    \midrule
    All features — baseline      & 1.6706 & — \\
    Remove $T_1$ (index 5)       & 1.6708 & $+0.0002$ \\
    Remove $T_2$ (index 6)       & 1.6705 & $-0.0001$ \\
    Remove Readout Error (idx 7) & 1.6151 & $-0.0555$ \\
    Remove CX Gate Error (idx 8) & 1.3771 & $-0.2935$ \\
    \bottomrule
  \end{tabular}
\end{table}

We consider effects with $|\Delta| < 0.001$ to be within numerical noise and not statistically meaningful.

\begin{figure}[htbp]
  \centering
  \includegraphics[width=\linewidth]{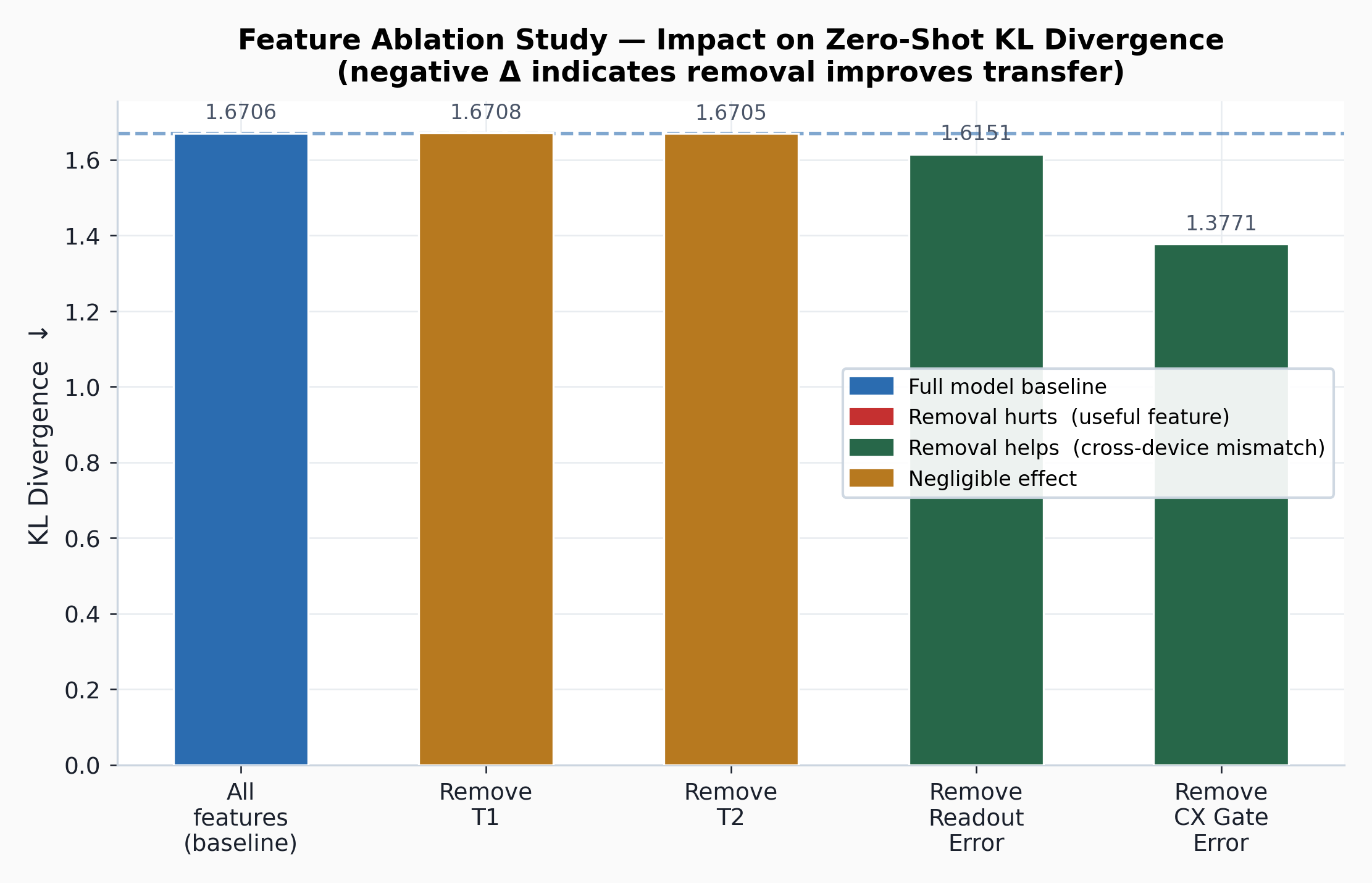}
  \caption{Feature ablation study. Bar heights show KL divergence when each calibration feature is zeroed out at inference time (model weights unchanged). Red bar: removing this feature increases KL (feature is informative under zero-shot transfer). Green bar: removing this feature decreases KL, indicating distribution mismatch rather than benefit—the model's learned association is miscalibrated for the target device's regime. Both readout error and CX gate error exhibit cross-device mismatch (negative $\Delta$), with CX gate error having the dominant effect.}
  \label{fig:ablation}
\end{figure}

\textbf{Finding 1 — CX gate error exhibits the strongest effect under ablation.} Removing CX gate error produces the largest KL reduction ($\Delta = -0.2935$), indicating that it introduces the most significant cross-device mismatch. The large distributional shift in CX error between devices ($+70.7\%$, Table~\ref{tab:calibration}) likely causes the model to misapply its learned CX-error-to-distortion mapping: the model was trained under \texttt{ibm\_fez}'s CX error regime and has learned to associate that regime's magnitude with specific distributional distortions. When applied to \texttt{ibm\_marrakesh}'s very different CX error magnitude, this learned association introduces systematic mismatch rather than meaningful correction. Under few-shot adaptation, the CX-sensitive layers would be updated to reflect the target device's regime, likely restoring or improving the feature's contribution.

\textbf{Finding 2 — Readout error exhibits weaker cross-device mismatch than CX gate error.} Readout error exhibits a weaker mismatch effect ($\Delta = -0.0555$), consistent with the 17.5\% difference in readout error between the two devices (Table~\ref{tab:calibration}) and with prior work on readout error mitigation~\cite{nation2023suppressing}: readout errors produce state-independent additive noise on measurement distributions, making them the most directly observable signature distinguishing the two devices.

\textbf{Finding 3 — $T_1$ and $T_2$ contribute minimally.} Both $T_1$ and $T_2$ ablations produce negligible changes ($|\Delta| < 0.001$). While coherence times differ between devices (Table~\ref{tab:calibration}), their mean-aggregated values do not encode the per-qubit spatial variation that determines local gate fidelity at the circuit level.

\section{Discussion}

\subsection{Why Does Few-Shot Adaptation Succeed?}

The residual architecture offers an explanation. The backbone network already captures patterns in how noise affects distributions (e.g. amplitude damping shifts mass towards the $|0\rangle$ state, readout errors symmetrically shift mass between bitstrings). This structure is common across devices, albeit to different degrees. To adapt to new devices, we only need to calibrate the calibration-dependent elements of this correction. The layer-freezing strategy reflects this: for small $K$, only the output head (which maps 64-dimensional backbone features to per-state corrections) is updated, limiting the adaptation to a $64 \times 32$ linear transformation. For $K = 20$, one additional hidden block is unfrozen. The replay buffer of 24 Backend A samples prevents catastrophic forgetting of this shared structural knowledge. This suggests that calibration features are not universally beneficial; their utility depends on alignment between source and target device regimes.

\subsection{Implications for Scalable Quantum Error Mitigation}

The 28.6\% KL improvement (relative to zero-shot) at $K = 20$—equivalently, 34.9\% recovery of the gap to in-domain performance—is achieved with only 20 real-hardware circuit executions: a negligible resource cost compared to full per-device data collection (85 circuits). This suggests a practical deployment strategy for quantum error mitigation at scale: train once on a well-characterized reference device, then adapt to new hardware with a small calibration set. The cost is proportional to $K$, not to the number of devices or circuit types.

\subsection{Comparison to Classical Error Mitigation}

Our approach differs structurally from ZNE and PEC. ZNE requires noise amplification circuits and extrapolation to the zero-noise limit—it does not use cross-circuit training data and cannot leverage a pre-trained prior. PEC requires quasi-probability sampling whose cost scales exponentially in circuit depth. In contrast, our approach requires $K$ real circuit executions for adaptation and then applies the corrected model with a single forward pass at test time. A direct quantitative comparison requires matching the circuit classes and evaluation metrics used in ZNE/PEC benchmarks, which we defer to future work.

\subsection{Practical Deployment Considerations}

One of the key applications for cross-device adaptation of noise is to support scalable deployment of quantum applications across different devices. In real-world cloud-based quantum computing systems, users may run jobs on multiple devices and the device calibrations may drift between sessions. Full retraining or large calibration efforts for each new device will cause high computational and operational costs. Hence, techniques allowing for fast adaptation with small datasets are critical for deployment.

The proposed approach enables an efficient deployment process. We first train a base model on a reference device with a large amount of data. This can be done offline and be periodically updated as new data is collected. To deploy on a different device, one takes a small set of $K$ circuits from the device of interest. This is followed by adaptation of the model using the few-shot adaptation protocol, resulting in a model for the new device. The adapted model can then be used to correct shots for any circuit, without the need for further calibration at runtime.

The adaptation procedure is not computationally expensive. Adaptation typically requires retraining only a small number of model parameters (such as the final layers or head) while keeping the other layers frozen. This speeds up training and reduces memory usage, making it feasible to incorporate within a quantum software pipeline. The inference step involves a single pass through a small neural network, leading to low overhead during inference compared to re-execution approaches.

Unlike classical error mitigation strategies based on repeated sampling or re-execution of circuits, the proposed approach amortizes the cost of adaptation over many subsequent executions of circuits. The cost of inference is proportional to the number of adaptation samples $K$ rather than the number of circuits that are executed in a deployment. This is especially attractive in a large workload scenario, where the same backend may be used to execute hundreds or thousands of circuits.

Moreover, the approach can easily be extended to multiple devices and time-varying scenarios. As more backends are seen, a common representation can be learned across devices, which may lead to better transfer learning capabilities in the future. The use of temporal training data may also enable adaptation over time as devices change, enabling the system to continue to perform satisfactorily without retraining.

These deployment scenarios indicate that few-shot cross-device adaptation is both an interesting research challenge and a practical approach to enhancing quality and scalability in near-term quantum devices.

\section{Limitations}

\begin{enumerate}
  \item \textbf{Dataset scale.} The dataset contains 170 samples across two devices and five random seeds. Conclusions about generalizability to the broader IBM fleet, to trapped-ion or photonic hardware, or to deeper and wider circuits should be drawn with caution.
  \item \textbf{Circuit scope.} Circuits are limited to 2--5 qubits and depth 2--8. Behavior under circuits approaching or exceeding coherence time limits (deeper circuits with thousands of gates) remains unstudied.
  \item \textbf{Mean-aggregated calibration features.} We use device-level mean values for $T_1$, $T_2$, readout error, and CX error. Per-qubit, per-gate, and topological calibration information would likely improve adaptation quality, particularly for spatially heterogeneous noise.
  \item \textbf{No temporal drift modeling.} Both device snapshots are single-point calibrations. IBM Quantum calibration data drifts substantially over hours to days~\cite{dasgupta2021stability}. A temporally-aware noise model is an important future direction.
  \item \textbf{Simple architecture.} We adopt a lightweight MLP architecture to isolate cross-device transfer effects without introducing confounding architectural complexity. Graph neural networks operating on the circuit's connectivity graph, or attention-based models over gate sequences, may better capture circuit-topology-dependent noise patterns. We do not include classical regression baselines (e.g., linear models or random forests); evaluating such baselines is an important direction for future work.
\end{enumerate}

\section{Conclusion}

We demonstrate that quantum noise is learnable, device-specific, and can be adapted with limited data. A residual neural network trained on IBM \texttt{ibm\_fez} exhibits a 5.5$\times$ increase in KL divergence under zero-shot transfer to \texttt{ibm\_marrakesh}, confirming that hardware-specific noise profiles do not generalize directly across devices. Few-shot fine-tuning with $K = 20$ target-device samples reduces this gap by 28.6\% relative to the zero-shot baseline, corresponding to a 34.9\% recovery toward in-domain performance, using a layer-selective adaptation strategy with replay to mitigate catastrophic forgetting.

Calibration drift analysis reveals qualitatively distinct noise profiles between the two devices, with the target device exhibiting longer coherence times but higher gate and readout error rates. Feature ablation indicates that CX gate error is the primary source of cross-device mismatch, while readout error contributes a smaller but consistent effect. These findings indicate that a modest number of calibration samples can be used to calibrate a pre-trained model to new devices.

Future directions include extending the method to larger sets of devices and deeper circuits, incorporating per-qubit and topology-specific features to the calibration, potentially using meta-learning algorithms like MAML~\cite{finn2017maml} to improve initialisation, and extending the method to capture temporal variations in noise. Overall, these results indicate that cross-device adaptation is a promising approach to scalable data-driven quantum error mitigation strategies.

\section*{Acknowledgments}

The authors acknowledge use of IBM Quantum services through the IBM Quantum Network. Circuit execution and calibration data were retrieved via the Qiskit IBM Runtime service. The views expressed are those of the authors and do not reflect the official policy of IBM or the IBM Quantum team.

\bibliographystyle{IEEEtran}
\bibliography{references}

\appendix

\section{Reproducibility Details}

All experiments use IBM Quantum real hardware accessed via Qiskit IBM Runtime. All 85 circuits per backend are submitted as a single batch job to minimize intra-session calibration drift. Each circuit is executed with 8192 shots. Calibration data is retrieved at the time of job submission via \texttt{backend.properties()}.

Model training: \texttt{seed = 42} for all random number generators (Python \texttt{random}, NumPy, PyTorch). Few-shot adaptation: averaged over seeds 0, 1, 2, 3, 4 for $K$-sample selection. All code is implemented in Python~3 using Qiskit, PyTorch, and NumPy. A complete reproducible pipeline is provided in the accompanying Jupyter notebook.

\end{document}